\documentclass[a4paper,11pt]{article}
\usepackage[utf8]{inputenc}
\usepackage[T1]{fontenc}
\usepackage{mathpazo}
\usepackage{amsmath,amssymb,amsthm}
\usepackage{csquotes}
\usepackage{color}
\usepackage{truthmakers}

\usepackage{multirow}
\usepackage{float}
\usepackage{url}
\usepackage{verbatim,enumitem}
\usepackage{blindtext}
\usepackage[hidelinks]{hyperref}
\usepackage{tikz-cd}
\usepackage{txfonts,pxfonts} 
\usepackage{pifont} 
\usepackage[bottom]{footmisc} 
\usepackage[all]{nowidow}

\setlist{itemsep=0pt} 
\setlist[1]{labelindent=\parindent} 
\usepackage[left=40mm,right=40mm,bottom=40mm]{geometry}
\usepackage[style=authoryear-comp, backend=biber, sortcites=ynt, sorting=nyt, doi=false, isbn=false, url=false]{biblatex}
\AtEveryBibitem{%
  \clearfield{number}
  \clearlist{language}
  \clearfield{month}
  \clearfield{day}
  \clearfield{endday}
  \clearfield{endmonth}
}

\newcommand{\gencite}[1]{\citeauthor{#1}'s (\citeyear{#1})}

\newcommand{\amrand}[2]
  {\leavevmode
   \marginpar
     [\raggedleft\scriptsize \leavevmode\normalcolor #1: #2]
     {\raggedright\scriptsize \leavevmode\normalcolor #1: #2}
  }

\newcommand{\tmm}[1]{|#1|_{\mathcal{M}}} 
\newcommand{\MM}{{\mathcal{M}}} 
\newcommand{\LLC}{\mathcal{L}^{\boxright}}

\title{Causal Modeling Semantics for Counterfactuals with Disjunctive Antecedents}
\author{Giuliano Rosella\footnote{Corresponding author. Center for Logic, Language and Cognition (LLC), Department of Philosophy and Education, Palazzo Nuovo, Via Sant'Ottavio 20, 10124 Torino, Italy. giuliano.rosella@unito.it} \and Jan Sprenger\footnote{Center for Logic, Language and Cognition (LLC), Department of Philosophy and Education, Palazzo Nuovo, Via Sant'Ottavio 20, 10124 Torino, Italy. jan.sprenger@unito.it}}

\begin{document}

\maketitle

\begin{abstract}
Causal Modeling Semantics \parencite[CMS, e.g.,][]{GallesPearl1998,Pearl2000,Halpern2000} is a powerful framework for evaluating counterfactuals whose antecedent is a conjunction of atomic formulas. We extend CMS to an evaluation of the probability of counterfactuals with disjunctive antecedents, and more generally, to counterfactuals whose antecedent is an arbitrary Boolean combination of atomic formulas. Our main idea is to assign a probability to a counterfactual $(A \vee B)\boxright C$ at a causal model $\mathcal{M}$ as a weighted average of the probability of $C$ in those submodels that \textit{truthmake} $A \vee B$ \parencite{Briggs2012,Fine2016,Fine2017}. The weights of the submodels are given by the inverse distance to the original  model  $\mathcal{M}$, based on a distance metric proposed by \textcite{EvaEtAl2019}. Apart from solving a major problem in the epistemology of  counterfactuals, our paper shows how work in semantics, causal inference and formal epistemology can be fruitfully combined. 
\end{abstract}



\pagebreak

\section{Introduction}\label{sec:intro}

How should we evaluate counterfactuals like ``if it had rained, the football match would have been cancelled''? There is an abundance of logical theories analyzing their truth conditions, such as the strict conditional analysis \parencite{Warmbrod1981JPL,Gillies2007Counterfactual} or premise semantics \parencite{Veltman1976,Kratzer1981Partition}. Much less theories, however, propose a unified treatment of the truth conditions and probability of counterfactuals. The principal contenders, and the ones to which we restrict our attention in this paper, are Stalnaker-Lewis similarity semantics (SLSS) and causal modeling semantics (CMS). 


 
Stalnaker-Lewis similarity semantics \parencite[SLSS:][]{Stalnaker1968,Lewis1973-Causation,Lewis1973-CF} is based on the idea that a counterfactual $A \boxright B$ is true at a possible world $w$ (i.e., a complete valuation of all sentences of the language) if and only if the consequent $B$ is true in the closest possible worlds where $A$ is true. Formally, for each sentence $A$, one defines a \textit{selection function} $f$ mapping a possible world $w$ to the set of $A$-worlds closest to $w$, denoted by $f_A(w)$. A counterfactual $A \boxright B$ is true if and only if $B$ holds at $f_A(w)$, and its probability is the cumulative probability of the possible worlds where it is true, i.e., $p(A \boxright B) = \sum_{w \models A \boxright B} p(w)$. 

As \textcite{Lewis1976} showed, this value does not correspond in general to the conditional probability $p(B|A)$. As an alternative algorithmic characterization of $p(A \boxright B)$, we may \textit{image} the probability distribution $p$ on the $A$-worlds: we assign the probability mass of the $\neg A$-worlds to the $A$-worlds that are closest to them, and evaluate $B$ relative to that distribution $p_A$. However, an equation of the type $p(A \boxright B) = p_A(B)$ holds only under specific assumptions on the selection function.\footnote{Specifically, it is required that the selection function  always chooses a unique closest $A$-world \parencite{Stalnaker1968,stalnaker1980defense}, but \textcite{Lewis1973-CF} argues that this assumption should be relaxed.} The probability of a counterfactual in SLSS is therefore still an open question \parencite[see also][]{Gardenfors1982,Guenther2022PCIP}. 

The dominant paradigm in computer science and formal epistemology, by contrast, is \textit{causal modeling semantics} (CMS): a counterfactual $A\boxright B$ is interpreted relative to a causal model $\MM$. It is true if an \textit{intervention} forcing the event $A$ in $\mathcal{M}$ also yields $B$, and false if this intervention does not yield $B$ \parencite[][]{Pearl2000,Pearl2017, Gibbard1978-GIBCAT-3}. This proposal, which relies on causal models as a graphical tool for reasoning and inference, is elaborated in \textcite{GallesPearl1998}. On this account, the ``probability of counterfactual statements'' \parencite[205]{Pearl2000} is interpreted as the probability that after an intervention on $A$ (written $do(A)$), $B$ will hold: $p(A \boxright B) = p_{do(A)}(B)$. 


The divergences and convergences of CMS and SLSS have been studied from various angles. \textcite[][72--73]{Pearl2000} shows that a particular type of imaging is equivalent to an intervention on $A$ that is represented by the $do$-operator (see also Section 6). It is agreed, however, that standard CMS and SLSS are different in at least one crucial respect: they assign truth conditions to different classes of counterfactuals. The SLSS framework assigns truth values---and probabilities---to counterfactuals $A \boxright B$ with \textit{arbitrary} antecedents, regardless of their logical complexity, since for any sentence $A$, the set of closest possible $A$-worlds is well-defined.\footnote{However, the interpretation and logical properties of counterfactuals with disjunctive antecedents  are the subject of substantive debate \parencite[e.g.,][]{Nute1975,Loewer1976,McKayInwagen1977}, and SLSS does not determine a canonical algorithm for calculating $p((A \vee B) \boxright C)$.} 

By contrast, Standard CMS, as developed in \textcite{GallesPearl1998}, cannot account for the truth conditions or probability of counterfactuals with disjunctive antecedents of the form $(A \vee B) \boxright C$, e.g., ``if it had rained \textit{or} there had been riots, the football match would have been cancelled''. The reason is that it is simply not clear which intervention corresponds to the logical \textit{disjunction} of two atomic interventions. In other words, while CMS has a  strong theoretical motivation and a history of successful applications, it has limited expressive power. To the extent that CMS aims at providing a semantics for natural language counterfactuals, this is a major limitation \parencite[see also][8]{Santorio2019Interventions}.  

Our paper closes the above gap: building on Briggs' \citeyear{Briggs2012} pioneering work on expanding CMS and ideas from truthmaker semantics \parencite{Fine2016,Fine2017}, we propose a CMS-based account for evaluating the probability of counterfactuals with disjunctive antecedents. Specifically, we propose to evaluate the probability of $(A \vee B) \boxright C$ as the weighted probability of $C$ in all submodels that \textit{truthmake} $A \vee B$. The relative weights of the submodels are determined by their distance to the original model, based on a metric developed in \textcite{EvaEtAl2019}. This procedure extends CMS to calculating the probability of counterfactuals with arbitrary Boolean compounds of atomic formulas in the antecedent. We also show that the predictions of our account are superior to the ones obtained in SLSS. 



The paper is structured as follows. In Section 2 and 3, respectively, we recapitulate the basics of causal modeling semantics and explain how truthmaker semantics can serve to establish a logic of counterfactuals. Section 4 introduces probabilistic causal models, Section 5 outlines our account and Section 6 compares it with the SLSS treatment of the probability of counterfactuals. Section 7 wraps up our results and suggests future work. 

\section{Causal Modeling Semantics (CMS)}\label{sec:cms}

This section recaps the causal modeling framework for the semantics of counterfactuals \parencite[CMS, e.g.,][]{GallesPearl1998,Pearl2000,Halpern2000}, as presented by \textcite{Briggs2012}. First, we need to introduce causal models, using a running example \parencite[simplified from][]{Pearl2000} that will accompany us throughout the paper. It involves four Boolean variables, whose values are represented by the numbers zero and one. 
\begin{quotation}\small
 A prisoner is condemned to death and led to the execution court. He stands in front of two soldiers, who will fire at the captain's signal. If at least one of the soldiers fires, the prisoner dies. The captain gives the signal ($C = 1$), the two soldiers fire ($X=1$, $Y=1$), and the prisoner dies ($D=1$). 
\end{quotation}
The main ingredients of this causal model are a set of variables $\mathcal{V}=\{C, X, Y, D\}$, and the set of structural equations that describe their causal dependencies: $\mathcal{S}=\{X=C, Y=C, D=max(X,Y)\}$. This means that the executioners fire if the captain gives the signal and the prisoner dies if one of the two executioners fires. The dependencies can also represented graphically, as in Figure \ref{fig:CM1} below. 

\begin{figure}[htb]
\begin{center}
\begin{tikzcd}[cells={nodes={draw=gray}}]
             & D                       &              \\
X \arrow[ru] &                         & Y \arrow[lu] \\
             & C \arrow[lu] \arrow[ru] &               
\end{tikzcd}
\end{center}
\caption{\small Causal graph for the prisoner execution story. $C$ stands for the captain (not) firing, $X, Y$ for the soldiers (not) shooting, $D$ for the prisoner dying/living.}\label{fig:CM1}
\end{figure}
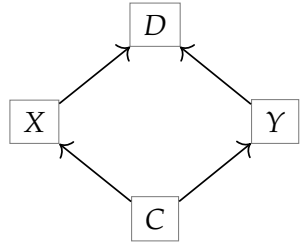

The \textit{parents} $PA(V)$ of a variable $V$ are simply the variables from which there is an arrow into $V$. For example, $C$ is the only parent of $X$ and $Y$, and $X$ and $Y$ are the parents of $D$. Structural equations describe the value of a variable as a function of the value of its parents. In general, a \textbf{causal model} $\mathcal{M}$ is a triple  $\mathcal{M}=\langle \mathcal{V}, \mathcal{S}, a\rangle$ where:
\begin{itemize}
    \item $\mathcal{V}$ is a non-empty finite set of variables $\mathcal{V}=\{V_1, V_2, ..., V_n\}$;
    \item $\mathcal{S}$ is a set of structural equations, where each element has the form $V = f_V(V_{i_1}, V_{i_2}, \ldots, V_{i_n})$ and $PA(V) = \{ V_{i_1}, \ldots, V_{i_n} \}$ (i.e., each structural equation defines the value of $V$ uniquely by the value of its parents; no cycles are allowed);
     \item $a: \mathcal{V} \rightarrow \mathcal{R}(\mathcal{V})$ is a function assigning an \textit{actual value} to each variable $V$, in a way that is consistent with the range of $V$ and the structural equations.
\end{itemize}

The last part, the assignment of actual values, is not necessarily required for making predictions with causal models, but it is crucial when we want to use them for counterfactual reasoning. 

Some additional terminology will be useful: when a variable $V_1$ is connected to another variable $V_2$ via a sequence of directed arrows from $V_1$ into $V_2$, we say that $V_2$ is a \emph{descendant} of $V_1$. For instance, in Figure \ref{fig:CM1}, $D$ is a descendant of $C$, $X$ and $Y$. As in \textcite{Briggs2012}, we will restrict our attention to  models not containing any loops, i.e., models where there is no sequence of arrows connecting a variable to itself. Moreover, in a causal model, we say that a variable is \emph{exogenous} when it has no parents (e.g., $C$ in Figure \ref{fig:CM1}) and \emph{endogenous} when it is not exogenous, so that its value can be determined by the value of other variables in the model (e.g., $X$, $Y$ and $D$ in Figure \ref{fig:CM1}). 

Now, we introduce the notion of an \textit{intervention} on a causal model. An atomic formula in our language has the form $V=v$, expressing the fact that the variable $V$ takes a certain value $v$. The intervention $do(V=v)$ on a causal model $\mathcal{M}$ breaks the dependency of $V$ on its parents via the structural equations (i.e., it eliminates all arrows into $V$) and assigns the value $V = v$ to it. The intervention generates a causal submodel $\mathcal{M'}$ where the formula $V = v$ is true and the structural equation $f_V$ is no longer part of the causal model: the variable $V$ now depends  on the intervention, but no longer depends on its parents. 

\begin{figure}[htb] 
\begin{center}
\begin{tikzcd}[cells={nodes={draw=gray}}]
             & D           &              \\
X \arrow[ru] &             & Y \arrow[lu] \\
             & C           &              
\end{tikzcd}
\end{center}
\caption{\small Causal graph for the prisoner execution story, where we intervene on $X$ and $Y$ and break the dependency on the captain's signal $C$.}\label{fig:CM2}
\end{figure}
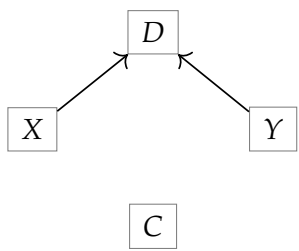

We can generalize this idea to conjunctions of interventions. For a causal model $\mathcal{M}=\langle \mathcal{V}, \mathcal{S}, a\rangle$, the intervention $do(V_1=v_1, V_2=v_2, \ldots, V_n=v_n)$ generates a \textbf{submodel} $\mathcal{M}'=\langle \mathcal{V}', \mathcal{S}', a' \rangle$ of $\mathcal{M}$ such that:

\begin{itemize}
    \item $\mathcal{V}'=\mathcal{V}$, i. e. $\mathcal{M}'$ has the same variables as $\mathcal{M}$;
    \item $\mathcal{S}'= \mathcal{S} \, \backslash \, \{ f_{V_1}, \ldots, f_{V_n} \}$; 
    \item $a': \mathcal{V} \, \backslash \, \{V_1, V_2, \ldots,V_n\} \to \mathcal{R}(\mathcal{V})$ assigns actual values to the variables not affected by the intervention, in line with the structural equations in $\mathcal{S}'$.
\end{itemize}
Conceptually, an intervention on a causal model manipulates some variables, forces them to take a certain value and breaks the causal mechanism between them and their parents. For an example, consider the causal model of the execution story depicted above; we want to know what would have happened if the two executioners had not fired $(X=0 \wedge Y=0)$. The answer is given by the intervention $do(X=0, Y=0)$ which would generate the model in Figure \ref{fig:CM2}.  

Our intervention has broken the causal mechanism that links $C$ to $X$ and $Y$, and we have forced $X$ and $Y$ to value zero. What happens to $D$ now? It continues to be determined by the structural equation $D=max(X, Y)$, but $X=0$ and $Y=0$ as a result of our intervention, hence $D=max(0, 0)=0$. And so the prisoner will live. 

Our intuitive counterfactual reasoning seems to run along these lines: in order to know \emph{what would have happened} to the prisoner \emph{had the executioners not fired}, we perform an intervention on the latter and see how it would have affected the prisoner, according to the known causal mechanisms, and without changing any facts that are causally independent of the executioners' actions. CMS explicates this line of thought in a mathematically precise way \parencite[e.g.,][205]{Pearl2000}. A counterfactual supposition in a causal model amounts to an external action on the model that enforces that supposition with a minimal change in the structure of the network (i.e., an intervention).

Specifically, a counterfactual sentence of the form $(A_1 \wedge A_2 \wedge ... \wedge A_n) \boxright B$ is true at a causal model $\mathcal{M}$ that contains $A_1, \ldots, A_n$ and $B$ as variables if and only if at the causal model $\mathcal{M}'$ generated by the intervention $do(A_1=1, A_2=1, \ldots, A_n=1)$ on $\mathcal{M}$, we also have $B=1$.\footnote{As before, we use $A_1 = 1$ for expressing that the Boolean variable $A_1$ takes the value ``true''.} For instance, the counterfactual \enquote{if the two executioners hadn't fired, then the prisoner would not have died} is true at the causal model of the execution story since, as we have seen above,  $D=0$ holds in the new submodel after performing the intervention $do(X=0, Y=0)$. 


Notice that an intervention of the form $do(A)$ is only defined when $A$ is an atomic formula or a conjunction of atomic formulas. This imposes a restriction on the class of counterfactuals that standard CMS can account for: only counterfactuals of the form $(A_1 \wedge A_2 \wedge ... \wedge A_n) \boxright B$ can assume a truth value. CMS does not provide truth conditions for counterfactuals with logically complex antecedents. For instance, we cannot say whether the counterfactual \enquote{if one of the two executioners hadn't fired, then the prisoner would not have died} ($(X=0 \vee Y=0) \boxright D=0$) is true or false at the causal model of the execution story. This limitation is due to the fact that the \emph{disjunctive intervention} $do(X=0 \vee Y=0)$ is not defined \parencite[see also][]{Pearl2017}. Intuitively, there is more than one possible realization of $do(X=0 \vee Y=0)$: we could manipulate $X$, $Y$, or both variables at the same time \parencite[compare][]{Sartorio2006,Briggs2012,Guenther2017}. Each of the three interventions $do(X=0)$, $do(Y=0)$ and $do(X=0, Y=0)$ would be a good candidate for an intervention that brings about the state ``$X=0$ or $Y=0$''. But their effects on $D=max(X, Y)$ differ. For the intervention $do(X=0)$ and $do(Y=0)$, the prisoner would still die (since the other soldier fires) but for the intervention $do(X=0, Y=0)$, he would live. Thus, if \emph{just one} executioner hadn't fired, the prisoner would have died anyway; if \emph{both} hadn't fired, he would live. So, in the end, standard CMS as presented in \textcite{GallesPearl1998} and \textcite{Pearl2000} does not provide a unique answer to the question of evaluating counterfactuals with disjunctive antecedents. This is arguably a disadvantage of CMS with respect to SLSS, where selection functions provide definite answers to the question of which are the relevant worlds for evaluating counterfactuals, and how the results need to be combined (e.g., Lewis demands that the consequent holds in all nearest possible worlds where the antecedent is true). In order to overcome this shortcoming, \textcite{Briggs2012} has proposed an extension of CMS that we present in the next section. 

\section{Truthmaker Semantics for Causal Modeling}\label{sec:tms}

Briggs' extension of CMS relies on truthmaker semantics (TMS), a semantic framework developed in a series of recent publications by Kit \textcite{Fine2016,Fine2017}. The idea underlying TMS is that of an \emph{exact truthmaker} of a sentence $A$, namely something in the world which is \emph{responsible} and \textit{wholly relevant} for the truth of $A$. One of the motivations behind truthmaker semantics is to be able to draw hyperintesional distinctions between propositions, i.e., to distinguish propositions that would be otherwise identical in the classical possible worlds framework, like $p$ and $p \vee (p \wedge q)$, or tautologies like $p \vee \neg p$ and $q \vee \neg q$. More precisely, the fundamental structure in TMS is that of a state space $\langle S, \sqsubseteq\rangle$ where $S$ is a non-empty set of states which stand for portions of reality (e.g., facts, events, individuals etc.), and $\sqsubseteq$ is a partial order over $S$ that can be understood as parthood relation between the elements in $S$. We can then define an operation $\sqcup$ of \textit{fusion} between states: given two states $s$ and $t$, their fusion of $s\sqcup t$ is the least upper bound of the set $\{s, t\}$. We can equip a state space with interpretation functions so as to define a relation of \emph{exact truthmaking} and \emph{exact falsemaking} between sentences and states so that those states can be truthmakers or falsemakers of formulas (for more details on TMS see for instance \textcite{Fine2017,KIT2019}). $s \Vdash A$ ($s \falsemakes A$) indicates that $s$ is an exact truthmaker (falsemaker) of $A$.  

\textcite{Briggs2012} shows how truthmaker semantics can expand the scope of CMS.  An intervention $do(A)$ is \textit{admissible} on a causal model $\mathcal{M}$ when it does not perform two inconsistent value assignments to the same variable, like $do(V_1=0 \wedge V_1=1)$. For a causal model $\mathcal{M}=\langle \mathcal{V}, \mathcal{S}, a\rangle$, we can define the \emph{set of submodels} of $\mathcal{M}$ generated by any intervention $do(A)$ as $S(\mathcal{M})=\langle S, \sqcup \rangle$ where


\begin{itemize}
    \item $S$ is the set of submodels of $\mathcal{M}$ generated by any admissible intervention $do(A)$;
    \item $\mathcal{M}[A]$ indicates the submodel generated by performing $do(A)$ on $\mathcal{M}$; 
    \item $\sqcup$ is an operation of \emph{fusion} among the models in $S$ defined by $\mathcal{M}[A] \sqcup \mathcal{M}[B] := \mathcal{M}[A \wedge B]$. 
\end{itemize}
In other words, the fusion of the two submodels $\mathcal{M}[A]$ and $\mathcal{M}[B]$, defined by the interventions $do(A)$ and $do(B)$, corresponds to the submodel defined by the fusion of the two interventions, where the fusion of two interventions is simply the intervention that encodes both, i.e., the conjunctive intervention of both of them. We assume that only logically consistent fusions are allowed. For instance, let $X$ be a variable in a model $\mathcal{M}$ which stands for the status of the light: $X=0$ means that the light is off, and $X=1$ means that the light is on. It is then impossible to fuse $\mathcal{M}[X=0]$ and $\mathcal{M}[X=1]$, because their fusion would yield a model where the light is both on and off, or in other words, the intervention $do(X=0 \wedge X=1)$ is not admissible.

Now, consider a language $\mathcal{L}$ where atomic formulas have the form $V=v$ and complex formulas are obtained from Boolean combinations of atomic formulas. For a model $\mathcal{M}$, consider its space of proper submodels $S(\mathcal{M})=\langle S, \sqcup\rangle$ where $\mathcal{M}\notin S$. We can inductively define relations of \textit{truthmaking} $\Vdash \ \subseteq S \times \mathcal{L}$ and and falsemaking $\falsemakes \, \subseteq S \times \mathcal{L}$ between any member $s$ of $S$ and formulas in the language as follows:
\[\begin{array}{lcl}
    s \Vdash V=v & \Leftrightarrow & s=\mathcal{M}[V=v] \\
    s \falsemakes V=v & \Leftrightarrow & s=\mathcal{M}[V=v'] \text{ for some } v \ne v'  \\
    s \Vdash \neg A & \Leftrightarrow & s \falsemakes A \\
    s \falsemakes \neg A & \leftrightarrow & s \Vdash A\\
    s \Vdash A \wedge B & \Leftrightarrow &  for\ some\ t, u \ (t \Vdash A, u \Vdash B \ and\ s = t\sqcup u)\\
    s \falsemakes A \wedge B & \Leftrightarrow &  s \falsemakes A, \ s \falsemakes B, \ or \ s\falsemakes A\vee B\\
    s \Vdash A \vee B & \Leftrightarrow &  s \Vdash A, \ s \Vdash B, \ or \ s\Vdash A\land B\\
    s \falsemakes A \vee B & \Leftrightarrow &  for\ some\ t, u \ (t \falsemakes A, u \falsemakes B \ and\ s = t\sqcup u)\\
\end{array}\]

\noindent
where $s \Vdash A$ means that $s$ \textit{truthmakes} (=is a truthmaker of) $A$. State $s$ is a truthmaker of $V=v$ if and only if it corresponds to the submodel defined by the intervention $do(V=v)$, and a falsemaker of $V=v$ if and only if it corresponds to the submodel defined by an intervention that sets $V$ to a value different from $v$. Since states in $S(\mathcal{M})$ can be identified with interventions, we can say, for simplicity, that an intervention $do(V_1=v_1, ..., V_n=v_n)$ on $\mathcal{M}$ truthmakes a formula $A$ if and only if $\mathcal{M}[V_1=v_1, ..., V_n=v_n]$ is a truthmaker of $A$. 

Evidently, $s$ falsemakes $A$ iff $s$ is a truthmaker of $\neg A$. State $s$ truthmakes a \textit{conjunction} of variable assignments iff it is the fusion of two states that truthmake the two individual assignments---in other words, iff $s$ is the causal submodel defined by the intervention that assigns the right values to both variables. Finally, $s$ is truthmaker of a \textit{disjunction} of variable assignments iff it truthmakes one of the two assignments, or its conjunction. This interpretation of truthmaking a disjunction is also at the center of Briggs' (and our own) proposal for expanding CMS. 

Consider a propositional language $\mathcal{L}$, which we extend to a language $\LLC$ with a simple, non-nested counterfactual operator: for any formulas $A, B \in \mathcal{L}$, let $A \boxright B \in  \LLC$.  We can now give inductively defined truth conditions for formulas of $\LLC$, including simple counterfactuals. 
\begin{description}
  \item[Truth Conditions for Formulas of $\LLC$ (Briggs)] A $\LLC$-formula is true at a causal model $\mathcal{M}=\langle \mathcal{V}, \mathcal{S}, a \rangle$ in the following conditions: 
\[\begin{array}{lcl}
    \mathcal{M} \vDash V=v & \Leftrightarrow & a(V)=v \\
    \mathcal{M} \vDash \neg A & \Leftrightarrow & \mathcal{M} \nvDash A  \\
    \mathcal{M} \vDash A \wedge B & \Leftrightarrow & \mathcal{M} \vDash A \ and \ \mathcal{M} \vDash B  \\
    \mathcal{M} \vDash A \vee B & \Leftrightarrow & \mathcal{M} \vDash A \ or \ \mathcal{M} \vDash B  \\
    \mathcal{M} \vDash A \boxright B & \Leftrightarrow & for \ every \ s \ in \ S(\mathcal{M}) \ such \ that\ s \Vdash A, s \vDash B  \\
\end{array}\]
\end{description}
Thus, a counterfactual $A \boxright B$ is true at a causal model $\mathcal{M}$ if and only if $B$ is true at all the members of $S(\mathcal{M})$ that truthmake $A$. Consider again the execution example and the counterfactual \enquote{if one of the two executioners had not fired, then the prisoner would not have died}. We can formalize this counterfactual as $(X=0 \vee Y=0) \boxright D=0$. The truthmakers of $X=0 \vee Y=0$ are the submodels $\mathcal{M}[X=0]$,  $\mathcal{M}[Y=0]$ and $\mathcal{M}[X=0 \wedge Y=0]$. The first two submodels validate $D=max(X,Y)=1$ since the second soldier is not affected by the intervention, and so $(X=0 \vee Y=0) \boxright D=0$ is false at $\mathcal{M}$. 

Briggs' extension of CMS allows us to assign a truth value to counterfactuals with disjunctive antecedents---and in fact, to counterfactuals with arbitrary Boolean compounds of atomic formulas in the antecedent. The main innovation to CMS consists in evaluating counterfactuals in the submodels that truthmake the antecedent. Implicit in Briggs' approach is a relevance principle for the truth conditions of counterfactuals, which we will also use later when defining their probability:
\begin{description}
  \item[Relevance Principle (Truth Conditions)] The truth value of a counterfactual $A \boxright B$ at a causal model $\mathcal{M}$ depends exclusively on the truth value of $B$ in the submodels $\mathcal{M}_1, \mathcal{M}_2, \ldots, \mathcal{M}_n$ generated by the interventions on the variables in $\mathcal{M}$ that truthmake $A$. 
\end{description}

Objections to truthmaker semantics will be dealt with in the discussion section: we now proceed to developing our proposal in the framework of probabilistic causal models.

\section{Probabilistic Causal Models}\label{sec:pcm}

In this section, we introduce probabilistic causal models and explain how CMS assigns a probability to counterfactuals. We will also see how the problem of the limited expressive power of CMS re-emerges at the probabilistic level: causal modeling semantics does not allow to assign a probability to counterfactuals with disjunctive antecedents.

A \textbf{probabilistic causal model} is a tuple $\mathcal{M}=\langle \mathcal{V}, \mathcal{G}, p\rangle$ where

\begin{itemize}
    \item $\mathcal{V}$ is a set of variables; 
    \item $\mathcal{G} \subset \mathcal{V} \times \mathcal{V}$ is a set of directed edges between the variables in $\mathcal{V}$, defining the parents and descendants of each variable; 
    \item $p$ is a probability distribution on $\mathcal{V}$ subject to the \textit{Markov condition}, that is, each variable $V$ is probabilistically independent of its non-descendants, conditional on its parents. 
\end{itemize}

 In a probabilistic causal model, the behavior of exogenous variables, and the dependencies of the endogenous variables on their parents, are described via a probability distribution. This differs from the non-probabilistic models of Section 2 whose variables are governed by structural equations.\footnote{The probabilistic nature of the models does not entail that the mechanism of dependence is intrinsically non-deterministic: for example, \textcite[][26]{Pearl2009} seems to favor the view that the non-deterministic dependencies of the variables are due the lack of knowledge about the underlying deterministic mechanism.} Consider again the execution scenario from Section 2 with the probability distribution $p$ described in Table \ref{tab:cpd}. Thanks to the Markov condition, it is sufficient to specify the probability of the exogenous variables, and the conditional probability of the endogenous variables, given the values of their parents. 

\begin{table}[htb]
\begin{center}
\begin{tabular}{ccccccccccc|cc|cc|}
\cline{1-2} \cline{4-6} \cline{8-10} \cline{12-15}
\multicolumn{2}{|c|}{C}                            & \multicolumn{1}{c|}{} & \multicolumn{1}{c|}{\multirow{2}{*}{C}} & \multicolumn{2}{c|}{X}         & \multicolumn{1}{c|}{} & \multicolumn{1}{c|}{\multirow{2}{*}{C}} & \multicolumn{2}{c|}{Y}         &  & \multirow{2}{*}{X} & \multirow{2}{*}{Y} & \multicolumn{2}{c|}{D} \\ \cline{1-2} \cline{5-6} \cline{9-10} \cline{14-15} 
\multicolumn{1}{|c|}{1} & \multicolumn{1}{c|}{0.5} & \multicolumn{1}{c|}{} & \multicolumn{1}{c|}{}                   & 1   & \multicolumn{1}{c|}{0}   & \multicolumn{1}{c|}{} & \multicolumn{1}{c|}{}                   & 1   & \multicolumn{1}{c|}{0}   &  &                    &                    & 0          & 1         \\ \cline{1-2} \cline{4-6} \cline{8-10} \cline{12-15} 
\multicolumn{1}{|c|}{0} & \multicolumn{1}{c|}{0.5} & \multicolumn{1}{c|}{} & \multicolumn{1}{c|}{1}                  & 0.9 & \multicolumn{1}{c|}{0.1} & \multicolumn{1}{c|}{} & \multicolumn{1}{c|}{1}                  & 0.9 & \multicolumn{1}{c|}{0.1} &  & 1                  & 0                  & 0.5        & 0.5       \\ \cline{1-2}
                        &                          & \multicolumn{1}{c|}{} & \multicolumn{1}{c|}{0}                  & 0.1 & \multicolumn{1}{c|}{0.9} & \multicolumn{1}{c|}{} & \multicolumn{1}{c|}{0}                  & 0.1 & \multicolumn{1}{c|}{0.9} &  & 0                  & 1                  & 0.5        & 0.5       \\ \cline{4-6} \cline{8-10}
                        &                          &                       &                                         &     &                          &                       &                                         &     &                          &  & 0                  & 0                  & 0.9        & 0.1       \\
                        &                          &                       &                                         &     &                          &                       &                                         &     &                          &  & 1                  & 1                  & 0.1        & 0.9       \\ \cline{12-15} 
\end{tabular}
\caption{\small Probability distribution for the variables in the execution example, as a function of the values of their parents. Intuitively, the table describes the dependencies among the variables; for instance we have that the value of $X$ will be 1 with $90\%$ probability if the value of $C$ is 1, i.e. $p(X=1|C=1)=0.9$. This means that it is almost certain that the executioner $X$ fires under the order of the captain, but there is a little chance ($10\%$) that $X$ might miss the shot, (for example, if the weapon jammed). Also, according to the table, it is almost certain that the prisoner dies if both the executioners fire, $p(D=1|X=1, Y=1)$, but there is a little chance ($10\%$) that he might survive. Of course, this probability distribution must be intended as a toy example.}\label{tab:cpd}
\end{center}
\end{table}

Analogously to the non-probabilistic case, probabilistic causal models provide an excellent tool for reasoning about counterfactuals. Again, the notion of an intervention is crucial. \textcite{Pearl2000} proposes that the probability of a counterfactual $A \boxright B$ at a probabilistic causal model $\mathcal{M}$, given a certain evidence $E$, amounts to the probability of $B$ in the submodel generated by the intervention $do(A)$ after updating on $E$, where $A$ is an atomic formula or a conjunction of atomic formulas. In other words, $p(A \boxright B|E) = p'_{do(A)}(B)$, where $p'(\cdot) = p(\cdot|E)$. This corresponds to the following procedure:
\begin{enumerate}
    \item Update the probability  $p(U=u)$ of each exogenous variable $U$ on the evidence $E$, via Bayesian conditionalization, to the new probability $p'(U'=u)=p(U=u|E)$, \textit{without changing the conditional dependencies among the variables}. This is because the evidence should not change the structure of the causal relationships between the variables: it just informs us which context we are likely to be in \parencite[see][33--38]{Pearl2000}. So $p'$ induces a new probability distribution on the (endogenous) variables, too. 
    \item Perform the intervention $do(A)$ on $\mathcal{M}$ to obtain a new submodel $\mathcal{M}'$ of $\mathcal{M}$; accordingly, change the probability distribution $p'$ so that variables involved in the intervention do not depend on their parents anymore. 
    \item Use the new submodel $\mathcal{M}'=\langle \mathcal{V}, \mathcal{G}', p'_{do(A)}\rangle$ with post-intervention graph $\mathcal{G}' \subseteq \mathcal{G}$ and probability distribution $p'_{do(A)}(\cdot)$ to calculate the probability of $B$ at $\mathcal{M}'$ (i.e., $p'_{do(A)}(B)$). 
\end{enumerate}


For example, consider the probabilistic execution model with the numbers from Table \ref{tab:cpd}. Assume that we have learned about the death of the prisoner, without knowing whether the captain has given the signal, or whether the executioners have fired. We have thus learnt the evidence $E = \{ D=1 \}$. By the procedure specified above, we need to update the probability of the \textit{exogenous variables}, i.e., $p'(C=1)=p(C=1|D=1)=0.82$, which induces a new probability distribution $p'$ on the endogenous variables.\footnote{Henceforth, unless otherwise stated, we will use $p'$ to refer to the probability distribution induced by $p'(C=1)=p(C=1|E)=0.82$.} Now, we want to compute the probability of $D=0$ under the counterfactual assumption that $X$ has not fired, $X=0$, corresponding to the probability to the counterfactual \enquote{if executioner $X$ hadn't fired, then the prisoner would not have died} ($X=0 \boxright D=0$). Following the above procedure, we should intervene by assigning value zero to $X$; this intervention $do(X=0)$ can be understood as an external action that forces the prisoner not to fire, for instance we sabotage $X$'s weapon. The action does not affect the probability of variables causally upstream of $X$: indeed our action is limited to $X$ and does not influence the behavior of $C$. Instead, it preempts the causal power of $C$ on $X$, and therefore we delete the arrow connecting $C$ to $X$. However, this intervention does affect the variables causally \textit{downstream} of $X$, imposing a new distribution on the model. Indeed, if we want how the prisoner is affected by this intervention, we need to calculate $p'_{do(X=0)}(D=0)$. Following the above procedure, we obtain that 
\begin{eqnarray*}
p'_{do(X=0)}(D=0) &=& \sum_{y, c \in \{0, 1\}}p(D=0|X=0, Y=y)\times p(Y=y|C=c)\times p(C=c|D=1) \\
&=& 0.598.
\end{eqnarray*}
In other words, it is 59.8\% probable that the prisoner would not have died under the counterfactual supposition that the executioner $X$ hadn't fired. This is, by the way, much less than the conditional probability $p'(D=0|X=0) = 0.752$ because \textit{updating on $X=0$} (with all other variables being unknown) would suggest an inference to the best explanation, i.e., that the captain did not give the signal. Hence, also the probability of $Y=0$ goes up sharply when we learn $X=0$, and so does the probability of $D=0$. 

Like deterministic CMS, the probabilistic framework does not account for the probability of counterfactuals with disjunctive antecedents since interventions are only defined for atomic formulas and their conjunctions. We will now develop a proposal that expands probabilistic CMS to arbitrary Boolean compounds of atomic formulas in the antecedent, similar to what Briggs has achieved for deterministic CMS. 

\section{CMS with Similarity Metrics}\label{sec:prop}

Suppose that we want to use probabilistic CMS in order to calculate the probability of a counterfactual with disjunctive antecedents. When we apply Pearl's procedure described in the previous section, steps 2 and 3 fail because the model generated by the intervention $do(X=0 \vee Y=0)$ is not well defined and consequently we cannot compute $p'(D=0)$.


A first step toward solving this problem is to impose a probabilistic version of the Relevance Principle from Section \ref{sec:tms}: 
\begin{description}
  \item[Relevance Principle (Probability)] The probability of a counterfactual $A \boxright B$ at a causal model $\mathcal{M}$ depends exclusively on the probability of $B$ in the submodels $\mathcal{M}_1, \mathcal{M}_2, \ldots, \mathcal{M}_n$ generated by the interventions on the variables in $\mathcal{M}$ that truthmake $A$. 
\end{description}
Thus, the probability of $(X = 0 \vee Y = 0) \boxright D=0$ depends exclusively on the probability of $D=0$ in the three submodels generated by $do(X=0)$, $do(Y=0)$ and $do(X=0 \wedge Y=0)$. See Table \ref{tab:CM3}. Step 2 is working now: performing the intervention $do(X=0 \vee Y=0)$ amounts to selecting three \emph{specific} submodels. However, step 3 is still problematic: it is not clear how the probabilities of $D=0$ in the three submodels should be combined. In fact, for $p'_{do(X=0)}(D=0) = p'_{do(Y=0)}(D=0) = 0.598$, whereas $p'_{do(X=0,Y=0)}(D=0)=0.9$. 
\begin{table}
\begin{center}
\begin{tikzcd}[cells={nodes={draw=gray}}]
             & D            &              &              & D            &              \\
X \arrow[ru] &              & Y=0 \arrow[lu] & X=0 \arrow[ru] &              & Y \arrow[lu] \\
             & C \arrow[lu] &              &              & C \arrow[ru] &              \\
             & do(Y=0)      &              &              & do(X=0)      &             
\end{tikzcd}
\end{center}

\begin{center}
\begin{tikzcd}[cells={nodes={draw=gray}}]
             & D                  &              \\
X=0 \arrow[ru] &                    & Y=0 \arrow[lu] \\
             & C                  &              \\
             & do(X=0 \wedge Y=0) &             
\end{tikzcd}
\end{center}
\caption{\small The three submodels that truthmake the sentence $X=0 \vee Y=0$ in the execution example, with the interventions used to generate them.}\label{tab:CM3}
\end{table}

It is clear that Briggs' solution for the truth conditions of a counterfactual with disjunctive antecedents will not help. There, the consequent needed to be true in \textit{all} states that truthmake the antecedent. \textcite[][152--154]{Briggs2012} recognizes that this is a \textit{choice}. The motivation is that there is no convincing argument for preferring a specific submodel. Moreover, also in Lewis' preferred version of SLSS, whenever there is a tie between the closest possible $A$-worlds, the counterfactual $A \boxright B$ is evaluated as true only if $B$ holds in \textit{all} of these worlds. While this is a reasonable choice in the context of a \textit{logic} of counterfactuals, we cannot transfer it to the \textit{probability} of counterfactuals where the output of the submodels are not Boolean values, but real numbers. We need to assign \textit{relative weights} to the truthmaking submodels, and this problem is specific to the probabilistic extension of Briggs' approach.


A natural requirement is that the values of $p'_s(B)$ in the relevant submodels indexed by $s$ should \textit{bound} the overall probability of the counterfactual $A \boxright B$ from above and below: 
\begin{description}
  \item[Convexity Principle] For the probability of a counterfactual $A \boxright B$ at a model $\mathcal{M}$, and the set of submodels $|A|_\mathcal{M}$ where we intervene on the variables in $\mathcal{M}$ as to truthmake $A$, 
  \begin{equation*}
\min (\{p_s(B): s \in |A|_\mathcal{M}\})\leq p(A \boxright B) \leq \max (\{p_s(B): s \in |A|_\mathcal{M}\})    
  \end{equation*}
  where $p_s$ denotes the probability distribution of the variables in submodel $s$, after updating on the available evidence and performing the truthmaking intervention. 
\end{description}
In other words, the probability of a counterfactual cannot be greater (smaller) than the maximum (minimum) probability of the consequent in the causal models that truthmake the antecedent  \parencite[see also][9]{Pearl2017}. 

The Convexity Principle still leaves space for a large class of weighting functions. A natural starting point is the \emph{straight average} of $p'(D=0)$ in the three submodels generated by $do(X=0 \vee Y=0)$. In this way, we would obtain $p'_{do(X=0\vee Y=0)}(D=0)=\frac{0.598+0.598+0.9}{3}=0.698$. However, straight averaging is at best a default assumption and devoid of a compelling motivation. An alternative is to make the relative weight of the three submodels generated by $do(X=0)$, $do(Y=0)$ and $do(X=0, Y=0)$ depend on their degree of \textit{similarity to the original model}. Once we have weights $\alpha_1, \alpha_2, \alpha_3$ for each of them, we can compute the post-intervention probability as 
\[
p'_{do(X=0\vee Y=0)}(D=0)=\alpha_1 \times 0.598 + \alpha_2 \times 0.598 + \alpha_3 \times 0.9.
\] 

The question is how to measure this degree of similarity. A possible answer comes from a recent work of \textcite{EvaEtAl2019} where the authors introduce two notions of similarity distance between causal models: \emph{evidential}  similarity distance, based on the shared probabilistic (in)depencies, and \emph{counterfactual} similarity distance, based on shared counterfactual dependencies. In what follows, we restrict our attention to the latter since probabilistic independencies can hide true causal and counterfactual dependencies.\footnote{In the causal modeling literature, this is known as failure of the Faithfulness Condition.} 
\begin{description}
\item[Counterfactual Dependence between Variables] A variable $V_2$ is counterfactually dependent on another variable $V_1$ when an intervention on $V_1$ affects the probability distribution of $V_2$, i.e., for some $v \in \mathcal{R}(\mathcal{V}_1)$, $p_{do(V_1 = v)}(V_2) \ne p(V_2)$.\footnote{For example, in the execution model, $D$ counterfactually depends on $X$, $Y$ and $C$; while $X$ and $Y$ counterfactually depends on $C$.}  
  \item[Counterfactual Similarity Distance (Eva et al., 2019)] Two (probabilistic) causal models $\mathcal{M}$ and $\mathcal{M}'$ are more or less similar to each other, the more counterfactual dependencies they agree on. Specifically, the counterfactual distance between $\mathcal{M}$ and $\mathcal{M}'$ is the absolute value of the difference of their counterfactual dependencies normalized by the total number of possible counterfactual dependencies:
  \[
  d(\mathcal{M}, \mathcal{M}')=\frac{|C_\mathcal{M}-C_{\mathcal{M}'}|}{N_C} \in [0,1].
  \] 
\end{description}
Recall that a variable $V_2$ is counterfactually dependent on another variable $V_1$ if we can go from $V_1$ to $V_2$ by following a sequence of arrows from $V_1$ to $V_2$: arrows represent the structural equations, i.e., the  \emph{mechanisms} or \emph{laws} that connect variables. Hence, if two models disagree on some counterfactual dependencies among the variables, they disagree on the \emph{mechanism} connecting those variables. So, intuitively, the more laws governing the original model are broken in $\mathcal{M}'$, the more counterfactual-distant from $\mathcal{M}$ a causal model $\mathcal{M}'$ is \parencite[see also][]{Lewis1973-Causation}.



There are two principled options for calculating the probability of counterfactuals. First, we could focus on the submodel that is most similar to $\mathcal{M}$ in the above metric, and neglect the contribution of the other submodels. This is feasible, but it would privilege a particular model and a specific way of truthmaking the antecedent. This is especially implausible when the truthmaking models have a similar distance to the original model and express qualitatively different ways of changing the mechanisms to make the antecedent true. 

Second, we could propose that the weight of each submodel $\mathcal{M}'$ should be inversely proportional to its distance to the original model $\mathcal{M}$, according to the above distance measure. This is our preferred approach since it takes into account all relevant submodels that truthmake the antecedent (and only them). 

For example, consider the execution story and the three submodels generated by $do(X=0)$, $do(Y=0)$ and $do(X=0 \wedge Y=0)$. The number of total pairwise counterfactual dependencies is $N_C = 12$; the original model $\mathcal{M}$ encodes $C_\mathcal{M} = 5$ counterfactual dependencies; each of the models generated by $do(X=0)$ and $do(Y=0)$ encodes $C_{\mathcal{M}'} = 4$ counterfactual dependencies and the model generated by $do(X=0 \wedge Y=0)$ encodes $C_{\mathcal{M}'} = 2$ counterfactual dependencies. Table \ref{tab:cfd} describes the counterfactual dependencies of the execution story and its submodels, where $V_1 \boxright V_2$ means that $V_2$ counterfactually depends on $V_1$:

\begin{center}
\begin{table}[H]
\begin{tabular}{c|c|c|c|c|}
\cline{2-5}
                              & $Original \ Model$      & $do(X=0)$               & $do(Y=0)$               & $do(X=0 \wedge Y=0)$    \\ \hline
\multicolumn{1}{|c|}{$C \boxright X$} & Yes                     & No                      & Yes                     & No                      \\ \hline
\multicolumn{1}{|c|}{$C \boxright Y$} & Yes                     & Yes                     & No                      & No                      \\ \hline
\multicolumn{1}{|c|}{$C \boxright D$} & Yes                     & Yes                     & Yes                     & No                      \\ \hline
\multicolumn{1}{|c|}{$X \boxright D$} & Yes                     & Yes                     & Yes                     & Yes                     \\ \hline
\multicolumn{1}{|c|}{$X \boxright Y$} & No                      & No                      & No                      & No                      \\ \hline
\multicolumn{1}{|c|}{$X \boxright C$} & No                      & No                      & No                      & No                      \\ \hline
\multicolumn{1}{|c|}{$Y \boxright D$} & Yes                     & Yes                     & Yes                     & Yes                     \\ \hline
\multicolumn{1}{|c|}{$Y \boxright X$} & No                      & No                      & No                      & No                      \\ \hline
\multicolumn{1}{|c|}{$Y \boxright C$}   & No                      & No                      & No                      & No                      \\ \hline
\multicolumn{1}{|c|}{$D \boxright X$} & No                      & No                      & No                      & No                      \\ \hline
\multicolumn{1}{|c|}{$D \boxright Y$} & No                      & No                      & No                      & No                      \\ \hline
\multicolumn{1}{|c|}{$D \boxright C$}  & \multicolumn{1}{|c|}{No} & \multicolumn{1}{|c|}{No} & \multicolumn{1}{|c|}{No} & \multicolumn{1}{|c|}{No} \\ \hline
\end{tabular}
\caption{\small Counterfactual Dependencies for the Execution Example.}\label{tab:cfd}
\end{table}
\end{center}

Call $\mathcal{M}$ the original execution model. By looking at the table we can deduce that
\begin{align*}
d(\mathcal{M}, \mathcal{M}[X=0]) &= \frac{1}{12} & d(\mathcal{M}, \mathcal{M}[Y=0]) &= \frac{1}{12} \\
d(\mathcal{M}, \mathcal{M}[X=0 \wedge Y=0]) & =\frac{3}{12}
\end{align*}
So, $\mathcal{M}[X=0]$ and $\mathcal{M}[Y=0]$ are equally similar to $\mathcal{M}$ and $\mathcal{M}[X=0 \wedge Y=0]$ is the most distant from $\mathcal{M}$. Hence, $\mathcal{M}[X=0 \wedge Y=0]$, which is the most distant submodel, will receive the least weight. Call $\tmm{A} = \{s | s \truthmakes A\}$ the set of truthmakers of $A$, i.e., the submodels generated by the intervention $do(A)$ on $\mathcal{M}$. In the model $\mathcal{M}$ of the execution story, 
\[
\tmm{X=0\vee Y=0} = \{\mathcal{M}[X=0], \mathcal{M}[X=0], \mathcal{M}[X=0\wedge Y=0] \}.
\] 
For $s \in \tmm{X=0\vee Y=0}$, we define its weight as 
\[
\alpha(s)= \frac{d(\mathcal{M}, s)^{-1}}{\sum_{t \in \tmm{X=0\vee Y=0}} d(\mathcal{M}, t)^{-1}},
\] 
following the rationale that the weight should be inversely proportional to the distance from the original model, normalized by the sum of all weights. 

By some computation, we get that 
\begin{align*}
\alpha(\mathcal{M}[X=0] &)=\alpha(\mathcal{M}[Y=0])=\frac{3}{7} & \alpha(\mathcal{M}[Y=0\wedge X=0]) &= \frac{1}{7}
\end{align*}
Applied to the execution story, we then find that 
\[
p'((X=0 \vee Y=0) \boxright D=0)=\frac{3}{7}\times 0.598 + \frac{3}{7}\times 0.598 + \frac{1}{7}\times 0.9\approx 0.64, 
\] 
in agreement with the Convexity Principle. We can generalize the weighting procedure as follows: for a causal model $\mathcal{M}$, for an arbitrary formula $A$ in $\mathcal{L}$, for $s \in |A|_\mathcal{M}$, 
\[
\alpha(s)= \frac{d(\mathcal{M}, s)^{-1}}{\sum_{t\in\tmm{A}} d(\mathcal{M}, t)^{-1}}. 
\]
Consequently, 
we calculate the probability of a counterfactual $A \boxright B$ with $\mathcal{L}$-sentences $A$ and $B$, relative to a causal model $\mathcal{M}$, as 
\begin{align}\label{eqn:cfprob}
p(A \boxright B) &= \sum_{s \in \tmm{A}} \alpha(s) \times p_s(B) \\
&= \sum_{s \in \tmm{A}} \frac{d(\mathcal{M}, s)^{-1}}{\sum_{t\in\tmm{A}} d(\mathcal{M}, t)^{-1}} \times p_s(B) \nonumber
\end{align}
Equation \eqref{eqn:cfprob} expresses our main idea in a nutshell: the probability of the counterfactual $p(A \boxright B)$ is the probability of the consequent $B$ in all submodels that truthmake the antecedent, weighted inversely by their similarity to the original model, where similarity is measured by the number of shared counterfactual dependencies. Our account thus synthesizes Causal Modeling Semantics with the Relevance Principle (=focusing on models that truthmake the antecedent, as in \textcite{Briggs2012}) and \gencite{EvaEtAl2019} proposal for measuring similarity between causal models. 


It is easy to see that our definition of the probability of a counterfactual with disjunctive antecedents extends to more complex sentences, too. Fine's truthmaker semantics indicates the truthmaking space states of all Boolean compunds of atomic sentences. Thus, for any sentence that we wish to take as the antecedent of a counterfactual, we simply determine the truthmaking states, the interventions on the causal model that correspond to them, and the corresponding counterfactual probabilities. Then we can use the Eva-Stern-Hartmann procedure for weighting the causal models that correspond to the truthmaking states. 

For example, if, for binary variables $A$ and $B$, our counterfactual is ``if $A = B$, then $C = 1$'' (with actual values $A=1$ and $B=0$), the antecedent has two truthmakers: the model generated by $do(A = 1, B = 1)$ and the one generated by $do(A = 0, B = 0)$. The two causal models obtained will then have the same weight according to our procedure, since the intervention affects the same variables and yields the same counterfactual dependencies. In other words, the probability of the counterfactual ``if $A = B$, then $C = 1$'' is simply the straight average of the probability of $C=1$ under the interventions $do(A=1, B=1)$ and $do(A=0, B=0)$.\footnote{Note that this also holds if it is \textit{actually} the case that $A=B=1$. Calculating the probability of the counterfactual does not privilege the actual values of variables; all that matters is whether the distance of the truthmaking models from the original model in terms of counterfactual dependencies.}

Taking stock, we have developed a procedure that goes beyond the achievements of \textcite{GallesPearl1998} and \textcite{Halpern2000}, who can calculate probabilities of counterfactuals, but only for antecedents representing a (conjunctive) set of interventions. On the other hand, \textcite{Briggs2012} has a general logic of counterfactuals, allowing for arbitary Boolean compounds as antecedents, but no extension to probabilistic reasoning. Our contribution provides a probabilistic counterpart of her logic motivated from the very same principles. 

\section{Back to Lewis: Comparison with Imaging}\label{sec:imaging}

In this section, we compare our account to the predictions of Stalnaker-Lewis similarity semantics (SLSS), and specifically, to \textit{imaging procedures} \parencite{Lewis1976,Gardenfors1982} for assigning a probability to a counterfactual with disjunctive antecedents. Imaging has been proposed not only in the logical analysis of counterfactuals, but also as an alternative to Bayesian conditionalization in the context of Causal Decision Theory \parencite{Joyce1999}, and so it needs to be taken seriously as a competitor to CMS. 

The basic ingredients of SLSS are a space of possible worlds $W$ together with a similarity order and a probability distribution $p$ on the elements of $W$. Possible worlds are complete valuations of the sentences of $\mathcal{L}$ and an ``$A$-world'' is a possible world where $A$ is true. The probability of a sentence $A \in \LLC$ is the cumulative  probability of the worlds where it is true, that is, $p(A)=\sum_{w \models A} p(w)$. For each $A \in \LLC$, we can moreover define a \textit{selection function} $f_A: W \to \mathcal{P}(W)$ that maps world $w$ to the $A$-worlds that are most similar to $w$, with the additional assumption that each world is most similar to itself (i.e., if $w \models A$, then $f_A(w) = \{w\}$).

This definition of probability does not yet allow for an algorithmic characterization of the probability of counterfactuals. Suppose therefore that the selection function $f_A$ always identifies a single closest $A$-world. Then we can define the probability distribution $p_A$ (``$p$ imaged on $A$'') as follows:
\begin{equation}\label{eqn:LewisIm} 
 p_A(w) := \sum_{v \in W} p(v) \times
     \begin{cases}
       1 & \text{if } f_A(v) = \{w\} \\
       0 & \text{otherwise}
     \end{cases}
\end{equation}
\textcite[310]{Lewis1976} shows that in this case, the probability of a counterfactual $A \boxright B$ is equal to the probability of $B$ after imaging on $A$: 
\[   
 p(A \boxright B) = p_A(B)= \sum_{w \models B} p_A(w) 
\]
According to \textcite[][311]{Lewis1976}, imaging is a ``minimal revision of the probability function to make the antecedent certain'', and this motivates why it could be the appropriate way of belief revision for evaluating a counterfactual. 

However, when the selection function does not identify unique possible worlds, and there can be ties between closest possible worlds, as argued by \textcite[77--83]{Lewis1973-CF}, we need to  generalize imaging beyond Equation \eqref{eqn:LewisIm}. \textcite{Guenther2022PCIP} shows that there are numerous ways of doing so, depending on how one distributes the mass of a $\neg$A-world $w$ among the selected worlds $f_A(w)$. For the purposes of counterfactual and causal reasoning, the following function proposed by \textcite{Gardenfors1982} is especially attractive: 
\begin{equation}\label{eqn:BMT}
 p_A(w)= \sum_{v \in W} p(v) \times
     \begin{cases}
       \frac{p(w)}{\sum_{w' \in f_A(v)} p(w')} & \text{if } w \in f_A(v)\\
       0 & \text{otherwise}
     \end{cases}
\end{equation} 
In this case, each world $w$ where $A$ is false transfers its probability mass to the closest worlds where $A$ is true, in proportion to the prior probability of these worlds. This type of imaging, which respects the prior probability ratio among the worlds that receive mass from $w$, is called \textit{Bayesianized imaging} by \textcite[]{Joyce1999}. Indeed, in the extreme case where $f_A(w) = \{v|v \models A\}$ if $w \not{in} A$ (i.e., all  $A$-worlds are selected), this form of imaging amounts to Bayesian conditionalization on $A$ \parencites[][73]{Pearl2000}[compare also Proposition 1 in][]{Guenther2022PCIP}.  

There is a deep connection between Bayesianized imaging and CMS. \textcite[]{Pearl2017} shows that the probability of a counterfactual $A \boxright B$, with $A = A_1 \wedge .... \wedge A_n$ being a conjunction of atomic formulas, can be characterized in two equivalent ways: either, in Causal Modeling Semantics, by  
\begin{equation}\label{eqn:Pearl1}
p(A \boxright B) := p_{do(A)}(B)
\end{equation}
or, \textit{when we count worlds with equal causal histories as equally similar}, and use the Bayesianized imaging function $p_A$ from Equation \eqref{eqn:BMT}, by 
\begin{equation}\label{eqn:Pearl2}
p(A \boxright B) := p_{A}(B)= \sum_{w \models B} p_{A}(w) 
\end{equation}
The first condition (``equal causal history'') means that the most similar $A$-worlds to a $\neg A$-world $w$ contain all and only those $A$-worlds that agree with $w$ on the value of the variables that cannot be affected by $do(A)$, i.e., the non-descendants of $A$. 

Pearl then shows that these two characterizations are equivalent, i.e.,
\begin{equation}\label{eqn:Pearl3}
 p_{A}(B) = p_{do(A)}(B).  
\end{equation}
In other words, the transformation defined by the $do$-operator can, for atomic interventions or their conjunctions, be interpreted as an imaging-type mass-transfer. This is a significant result showing that Bayesianized imaging and CMS agree for a large class of interventions. This result also motivates why we put Bayesianized imaging (as opposed to, e.g., equal weights imaging) at the center of the comparison of our own proposal with SLSS.

\begin{table}[htb]
\centering
		\begin{tabular}{|c|c|c|c|c|c|c|}
	\hline
	Worlds & \multicolumn{4}{c|}{\centering{Values}} & \multicolumn{2}{c|}{Closest worlds for imaging $w_i$ on $X=0 \vee Y=0$}                     \\ \cline{2-5} 
	 & $C$ & $X$ & $Y$ & $D$ & Option 1: $f_1(w_i) = \ldots$        & Option 2: $f_2(w_i) = \ldots$ \\ \hline
		$w_1$ & 1 & 1 & 1 & 1 & $\{w_3, w_4, w_7, w_8\}$ & $\{w_3, w_4, w_5, w_6, w_7, w_8\}$ \\ \hline
		$w_2$ & 1 & 1 & 1 & 0 & $\{w_3, w_4, w_7, w_8\}$ & $\{w_3, w_4, w_5, w_6, w_7, w_8\}$ \\ \hline
		$w_3$ & 1 & 1 & 0 & 1 & $\{w_3\}$ & $\{w_3\}$                 \\ \hline
  	$w_4$ & 1 & 1 & 0 & 0	& $\{w_4\}$ & $\{w_4\}$                              \\ \hline
		$w_5$ & 1 & 0 & 1 & 1	& $\{w_5\}$ & $\{w_5\}$                         \\ \hline
  	$w_6$ & 1 & 0 & 1 & 0	& $\{w_6\}$ & $\{w_6\}$                             \\ \hline
		$w_7$ & 1 & 0 & 0 & 1	& $\{w_7\}$ & $\{w_7\}$                            \\ \hline 
	  $w_8$ & 1 & 0 & 0 & 0	& $\{w_8\}$ & $\{w_8\}$                             \\ \hline
		$w_9$ & 0 & 1 & 1 & 1	& $\{w_{11}, w_{12}, w_{15}, w_{16}\}$ & $\{w_{11}, w_{12}, w_{13}, w_{14} w_{15}, w_{16}\}$                             \\ \hline
		$w_{10}$ & 0 & 1 & 1 & 0	& $\{w_{11}, w_{12}, w_{15}, w_{16}\}$ & $\{w_{11}, w_{12}, w_{13}, w_{14} w_{15}, w_{16}\}$                             \\ \hline
		$w_{11}$ & 0 & 1 & 0 & 1	& $\{w_{11}\}$ & $\{w_{11}\}$                             \\ \hline
		$w_{12}$ & 0 & 1 & 0 & 0	& $\{w_{12}\}$ & $\{w_{12}\}$                             \\ \hline
		$w_{13}$ & 0 & 0 & 1 & 1	& $\{w_{13}\}$ & $\{w_{13}\}$                             \\ \hline
		$w_{14}$ & 0 & 0 & 1 & 0	& $\{w_{14}\}$ & $\{w_{14}\}$                             \\ \hline
		$w_{15}$ & 0 & 0 & 0 & 1	& $\{w_{15}\}$ & $\{w_{15}\}$                             \\ \hline
		$w_{16}$ & 0 & 0 & 0 & 0	& $\{w_{16}\}$ & $\{w_{16}\}$                             \\ \hline
		\end{tabular}
	\caption{\small Two plausible selection functions $f_1$ and $f_2$ in the execution example with disjunctive interventions. The two selection functions correspond to two different ways of identifying, for any $w_i \in W$, the closest possible world where $X=0 \vee Y=0$ holds.}\label{tab:mswe1}
	\end{table}

We now extend Bayesianized imaging to the probability of counterfactuals with disjunctive antecedents. Consider the execution model again. We associate a possible world $w$ to each possible realization of the binary variables $C$, $X$, $Y$, $D$; so there are 16 possible worlds in total. The probability of each of them is simply the joint probability of the realizations of the variables in that possible world, respecting the conditional independence relations imposed by model $\MM$ and the Causal Markov Condition. For modeling Bayesianized imaging on a sentence $A$, we develop a three-step procedure analogous to the one recommended by CMS: 
\begin{enumerate}
  \item Update the prior probability of the exogenous variables $U$ on the observed evidence $E$ from $p(U=u)$ to the posterior probability $p'(U=u) = p(U=u|E)$. For all endogenous variables, their conditional probability distribution continues to be given by the probabilistic causal model $\MM$. 
  \item Transfer the mass of the $\neg A$-worlds to the closest possible $A$-worlds (chosen by the selection function $f$), weighted by the posterior probability of the latter. This will yield the probability function $p'_A(\cdot)$. 
  \item Calculate the probability of any sentence $B$ as $p'_A(B)$. 
\end{enumerate}

In the execution model, only $C$ is an exogenous variable and this means that the joint posterior distribution after the first step of the above procedure will look as follows:
\[
p'(C, X, Y, D ) = p'(C)\times p(X|C) \times p(Y|C) \times p(D| X, Y) 
\]
Now we proceed to the second step and image $p'$ on $(X=0 \vee Y=0)$. This means that four worlds will have weight zero in $p'_{X=0 \vee Y=0}$: $w_1$, $w_2$, $w_9$ and $w_{10}$ in Table \ref{tab:mswe1}. The question is how their weight should be distributed to the rest; and this depends on what are the closest neighbors to these possible worlds. 


	
The first conceptual obstacle in defining a similarity order is to decide which variables are \textit{not} affected by $do(X=0 \vee Y=0)$. Again, we translate the problem into Causal Modeling Semantics. According to \textcite{Briggs2012}, the disjunctive intervention  $do(X=0 \vee Y=0)$ can be regarded as encoding three different interventions, $do(X=0)$, $do(Y=0)$, and $do(X=0 \wedge Y=0)$. The closest worlds to $w_1$ for the first intervention are $w_7$ and $w_8$, for the second, they are $w_3$ and $w_4$, and for the third, $w_5$ and $w_6$. Depending on how seriously we consider the option of intervening on \textit{both} variables as a way of expressing $do(X=0 \vee Y=0)$, this gives us two options for the most similar worlds to $w_1$: $\{w_3, w_4, w_7, w_8\}$ or $\{w_3, w_4, w_5, w_6, w_7, w_8\}$. And vice versa for the other worlds whose weight needs to be cancelled. Both options are represented in the rightmost columns of Table \ref{tab:mswe1}.\footnote{A potential third option that also takes into account the value of $D$, i.e., $f(w_1) =   \{w_3, w_7\}$, does not yield qualitatively different results.}

However, if we calculate the probability of the counterfactual $(X=0 \vee Y=0) \boxright D=0$, after having learnt the evidence $D=1$, the result of Bayesianized imaging will, for either of these similarity orders, differ from our proposal. For Option 1, we obtain $p'_{X=0 \vee Y=0}(D=0)\approx 0.56$, and for Option 2, we obtain $p'_{X=0 \vee Y=0}(D=0)\approx 0.57$.\footnote{Alessandro Zangrandi's GitHub \url{https://github.com/zazangra/lewis_imaging} offers a  Python program to perform Bayesianized imaging on a causal model.} This is arguably a bad prediction since it violates the plausible Convexity Principle: the probability of the counterfactual should be bounded from above and below by the (maximal and minimal) probability of the consequent in the causal submodels that truthmake the antecedent. To recall: 
\begin{align*}
 p'(X=0 \boxright D=0)&=0.598  &  p'((X=0 \wedge Y=0) \boxright D=0) &=0.9 \\
 p'(Y=0 \boxright D=0)&=0.598
\end{align*}

To the extent that the Convexity Principle is plausible and compelling, we should reject any procedure that violates this constraint. Why should the probability of the counterfactual be above or below the probability of the consequent in all relevant submodels? It is simply paradoxical that the death of the prisoner, $D=1$, is more probable under the hypothetical assumption that \emph{at least one} of the two executioners did not fire ($p'_{X=0 \vee Y=0}(D=0)\approx 0.56$) than under the assumption that \emph{only one} did not fire ($p'_{X=0}(D=0)=0.598$).


Primarily, the failure of Convexity in imaging is due to the fact that there is no systematic connection between $p'_{X=0} (D=0)$ and $p'_{X=0 \vee Y=0} (D=0)$, like in our own proposal. For instance, when imaging on $X=0$, part of the mass of $w_3$ is transferred to $w_5$, whose probability mass makes a contribution to $p'_{X=0}(D=0)$, but not to $p'_{X=0\vee Y=0}(D=0)$ (in Option 1). This explains why the latter probability falls below $p'_{X=0} (D=0)$, i.e., below the bounds resulting from the Convexity Principle. In other words, the violation of the Convexity Principle is due to the fact that Bayesianized imaging does not respect the Relevance Principle: the possible worlds do not contain any information about the causal structure of the model.

Of course, generalized imaging offers an entire universe of different mass transfer functions.  So we do not exclude that the imaging theorist can find a function that complies with the Convexity Principle.\footnote{Equal weights imaging, a possible alternative, respects the Convexity Principle because it trivializes the problem: imaging on $X=0$, $Y=0$, $X=0 \wedge Y=0$ and $X=0 \vee Y=0$ all yield the same probability $p'(D=0) = 0.5615$. This is obviously an unacceptable result.} However, this must come at the price of choosing a procedure that deviates systematically from CMS for (conjunctions of) atomic interventions. What the imaging theorist \textit{cannot} have is a probability mass transfer function that agrees in regular circumstances with CMS, and that satisfies at the same time the Convexity Principle when applied to more complex interventions. Indeed, \textcite[][6--7]{Pearl2017} explicitly advises caution when applying imaging to disjunctive interventions, such as the ones that we discussed in this paper. Hence, we conclude that the combination of SLSS and imaging has not yet delivered a convincing response to the problem of evaluating the probability of counterfactuals with disjunctive antecedents.

\section{Conclusions}\label{sec:ccl}

The present paper extends Causal Modeling Semantics to the evaluation of the probability of counterfactuals with disjunctive antecedents, and more generally, to any counterfactuals whose antecedents are truth-functional compounds of atomic sentences. To the best of our knowledge, no other proposal has been advanced in the literature to achieve this goal. Our approach is very natural combines three well-established ideas: (1) Briggs' characterization of disjunctive interventions relying on truthmaking causal submodels; (2) weighting the contributions of these submodels according to their similarity with the original world; (3) Eva et al.'s definition of a similarity metric between causal models by counting shared counterfactual dependencies. 

As an alternative to our approach, one can assign probabilities to counterfactuals with disjunctive antecedents by imaging mass transfers, and Bayesianized imaging in particular. However, this option does not return plausible predictions about the probability of counterfactuals. What is more, it violates intuitive requirements such as the Convexity Principle and the Relevance Principle. 
 
It could be objected that we have not motivated the use of truthmaker semantics properly, and that we could also use, as an alternative, \textit{complete} value assignments to  the variables in the antecedent, e.g., $\mathcal{M}[X=0, Y=1]$,  $\mathcal{M}[X=1,Y=0]$ and $\mathcal{M}[X=0,Y=0]$. Our response is twofold: First, our contributions in Section 5 and 6 would still stand since nothing specific depends on the choice of truthmaker semantics in weighting the contributions of the submodels. If the readers prefer a different set of relevant submodels, they could still follow our similarity-based weighting procedure for assigning a probability to counterfactuals with disjunctive antecedents. This is our dialectical point. The substantial point is that the truthmaking submodels of a disjunction are consistent with each other under state fusion; this is \textit{not} true of the above alternative proposal. 

Another open question is whether our work is really an explication of the ``probability of counterfactuals''. CMS reads this term as the probability of a causal effect, given a minimal intervention, but it is not clear whether this really corresponds to the probability (plausibility, assertability) of a counterfactual sentence. Experiments in linguistics would be required to confirm the adequacy of the CMS interpretation, and our principle that the probability of a counterfactual $A \boxright B$ should be bounded from above and below by the \emph{best} and \emph{worst} scenarios for $B$ that we could imagine when supposing $A$. Another future application of our work is to shed new lights on the notion of \emph{disjunctive causes} introduced by \textcite{Sartorio2006}. Finally, we should spell out the implications of our findings for  premise semantics and their relationship to causal modeling semantics \parencite[][]{Kaufmann2013,Santorio2019Interventions}. 


 
\subsection*{Acknowledgments}
We would like to thank two anonymous reviewers of this journal for helpful comments. This work was supported by the Horizon2020 program of the European Commission through ERC Starting Grant No. 640638; and the Italian Ministry of University and Research through PRIN grant ``From Models to Decisions''.
 

\printbibliography

@InCollection{stalnaker1980defense,
  author    = {Stalnaker, Robert},
  booktitle = {Ifs},
  publisher = {Springer},
  title     = {{A Defense of Conditional Excluded Middle}},
  year      = {1980},
  editor    = {William Harper},
  pages     = {87--104},
}

@incollection{Stalnaker1968,
	Address = {Oxford},
	Author = {Stalnaker, Robert},
	Booktitle = {Studies in Logical Theory: American Philosophical Quarterly Monograph Series, No. 2},
	Publisher = {Blackwell},
	Title = {{A Theory of Conditionals}},
	Year = {1968}}

@article{Lewis1976,
	Author = {David Lewis},
	Journal = {Philosophical Review},
	Number = {3},
	Pages = {297--315},
	Publisher = {Duke University Press},
	Title = {Probabilities of Conditionals and Conditional Probabilities},
	Volume = {85},
	Year = {1976}}

@book{Pearl2009,
	Address = {Cambridge},
	Author = {Pearl, Judea},
	Edition = {2nd},
	Publisher = {Cambridge University Press},
	Title = {{Causality}},
	Year = {2009}}

@Book{Guenther2022PCIP,
  author = {Günther, Mario},
  title  = {{Probabilities of Conditionals and Imaged Probabilities}},
  year   = {2022},
}

@Article{Briggs2012,
  author  = {Briggs, R.A.},
  journal = {Philosophical Studies},
  title   = {Interventionist Counterfactuals},
  year    = {2012},
  number  = {1},
  pages   = {139--166},
  volume  = {160},
  date    = {2012},
}

@Book{Pearl2000,
  title     = {{Causality}},
  publisher = {Cambridge University Press},
  year      = {2000},
  author    = {Pearl, Judea},
  address   = {Cambridge},
}

@Article{GallesPearl1998,
  author  = {David Galles and Judea Pearl},
  journal = {Foundations of Science},
  title   = {An Axiomatic Characterization of Causal Counterfactuals},
  year    = {1998},
  pages   = {151--182},
  volume  = {3},
  date    = {1998},
}

@Article{Pearl2017,
  author  = {Pearl, Judea},
  title   = {{Physical and Metaphysical Counterfactuals: Evaluating Disjunctive Actions}},
  journal = {{Journal of Causal Inference}},
  year    = {2017},
  volume  = {5},
  number  = {2},
  pages   = {1--10},
  date    = {2017},
}

@Article{Warmbrod1981JPL,
  author  = {Warmbrod, Ken},
  journal = {{Journal of Philosophical Logic}},
  title   = {{Counterfactuals and Substitution of Equivalent Antecedents}},
  year    = {1981},
  pages   = {267--289},
  volume  = {10},
}

@Article{Gillies2007Counterfactual,
  author  = {Gillies, Anthony},
  journal = {{Linguistics and Philosophy}},
  title   = {Counterfactual Scorekeeping},
  year    = {2007},
  pages   = {329--360},
  volume  = {30},
}

@Article{Kratzer1981Partition,
  author    = {Angelika Kratzer},
  title     = {Partition and Revision: The Semantics of Counterfactuals},
  journal   = {Journal of Philosophical Logic},
  year      = {1981},
  volume    = {10},
  number    = {2},
  pages     = {201--216},
  doi       = {10.1007/bf00248849},
  publisher = {Springer},
}

@InCollection{Gibbard1978-GIBCAT-3,
  author    = {Allan Gibbard and William L. Harper},
  title     = {Counterfactuals and Two Kinds of Expected Utility},
  booktitle = {Ifs},
  publisher = {D. Reidel},
  year      = {1976},
  editor    = {W. L. Harper and R . Stalnaker and G. Pearce},
  pages     = {153--169},
}

@Article{Lewis1973-Causation,
  author    = {David Lewis},
  title     = {Causation},
  journal   = {Journal of Philosophy},
  year      = {1973},
  volume    = {70},
  number    = {17},
  pages     = {556--567},
  publisher = {Oxford Up},
}

@Book{Lewis1973-CF,
  title         = {Counterfactuals},
  publisher     = {Blackwell},
  year          = {1973},
  author        = {Lewis, David},
  address       = {Oxford},
}

@Article{Halpern2000,
  author  = {Halpern, Joseph Y.},
  title   = {Axiomatizing Causal Reasoning},
  journal = {Journal of Artificial Intelligence Research},
  year    = {2000},
  volume  = {12},
  pages   = {317--337},
}

@Article{EvaEtAl2019,
  author  = {Eva, Benjamin and Stern, Reuben and Hartmann, Stephan},
  title   = {The Similarity of Causal Structure},
  journal = {Philosophy of Science},
  year    = {2019},
  volume  = {86},
  number  = {5},
  pages   = {821--835},
  date    = {2019},
}

@Article{Fine2016,
  author  = {Fine, Kit},
  title   = {Angellic Content},
  journal = {Journal of Philosophical Logic},
  year    = {2016},
  volume  = {45},
  number  = {2},
  pages   = {199--226},
  date    = {2016},
}

@InCollection{Fine2017,
  author    = {Fine, Kit},
  title     = {{Truthmaker Semantics}},
  booktitle = {{A Companion to the Philosophy of Language}},
  publisher = {Wiley},
  year      = {2017},
  editor    = {Bob Hale and Crispin Wright and Alexander Miller},
  pages     = {556--577},
  address   = {New York},
  date      = {2017},
}

@Book{Joyce1999,
  title     = {{Foundations of Causal Decision Theory}},
  publisher = {Cambridge University Press},
  year      = {1999},
  author    = {Joyce, James},
  address   = {Cambridge},
}

@Article{Sartorio2006,
  author  = {Sartorio, Carolina},
  title   = {{Disjunctive Causes}},
  journal = {{Journal of Philosophy}},
  year    = {2006},
  volume  = {103},
  number  = {10},
  pages   = {521--538},
  date    = {2006},
}

@Article{Gardenfors1982,
  author  = {Gärdenfors, Peter},
  title   = {{Imaging and Conditionalization}},
  journal = {{Journal of Philosophy}},
  year    = {1982},
  volume  = {79},
  pages   = {747--760},
}

@Article{Loewer1976,
  author  = {Loewer, Barry},
  title   = {Counterfactuals with Disjunctive Antecedents},
  journal = {Journal of Philosophy},
  year    = {1976},
  volume  = {73},
  number  = {16},
  pages   = {531--537},
}

@Article{McKayInwagen1977,
  author  = {McKay, Thomas and Van Inwagen, Peter},
  title   = {{Counterfactuals with Disjunctive Antecedents}},
  journal = {Philosophical Studies},
  year    = {1977},
  volume  = {31},
  pages   = {353--356},
}

@Article{Nute1975,
  author  = {Nute, Donald},
  title   = {{Counterfactuals and the Similarity of Worlds}},
  journal = {{Journal of Philosophy}},
  year    = {1975},
  volume  = {72},
  number  = {21},
  pages   = {773--778},
}

@InProceedings{Guenther2017,
  author    = {Günther, Mario},
  title     = {{Disjunctive Antecedents for Causal Models}},
  booktitle = {Proceedings of the 21st Amsterdam Colloquium},
  year      = {2017},
}

@Article{KIT2019,
  author    = {Kit Fine and Mark Jago},
  journal   = {The Review of Symbolic Logic},
  title     = {Logic for Exact Entailment},
  year      = {2019},
  month     = {feb},
  number    = {3},
  pages     = {536--556},
  volume    = {12},
  publisher = {Cambridge University Press ({CUP})},
}

@Article{Kaufmann2013,
  author  = {Stefan Kaufmann},
  title   = {Causal Premise Semantics},
  journal = {Cognitive Science},
  year    = {2013},
  volume  = {37},
  number  = {6},
  pages   = {1136--1170},
  doi     = {10.1111/cogs.12063},
}

@Article{Santorio2019Interventions,
  author  = {Paolo Santorio},
  title   = {Interventions in Premise Semantics},
  journal = {Philosophers' Imprint},
  year    = {2019},
  volume  = {19},
}

@InCollection{Veltman1976,
  author    = {Frank Veltman},
  title     = {Prejudices, presuppositions, and the theory of counterfactuals},
  booktitle = {Amsterdam Papers in Formal Grammar. Proceedings of the 1st Amsterdam Colloquium},
  publisher = {University of Amsterdam},
  year      = {1976},
  editor    = {Jeroen Groenendijk and Martin Stokhof},
  pages     = {248--281},
}

\end{document}